\documentclass[prl,aps,twocolumn,showpacs,preprintnumbers, superscriptaddress, amsmath,amssymb]{revtex4-1}
\usepackage{comment}
\usepackage{xcolor} 
\usepackage{soul}
\usepackage{microtype}
\usepackage{amsmath}
\usepackage{dcolumn}
\usepackage{amssymb}
\usepackage{amsthm}
\usepackage{mathrsfs}
\usepackage{graphicx}
\usepackage{verbatim}
\usepackage{bm}
\usepackage[colorlinks, linkcolor=blue, anchorcolor=blue, urlcolor=blue, citecolor=blue]{hyperref}
\usepackage{bm}
\usepackage{times}
\usepackage{lipsum}

\usepackage[all]{xy}
\usepackage{ds font}

\newcommand{\be}{\begin{eqnarray}}
\newcommand{\ee}{\end{eqnarray}}

\newcommand{\ra}{\rangle}
\newcommand{\rar}{\rightarrow}

\begin{document}
	\title{ Restoring  Kibble-Zurek Scaling and Defect Freezing in Non-Hermitian Systems under Biorthogonal Framework}
	\author{Menghua Deng}
  \author{Wei Li}
  \author{Kangyi Hu}
	\author{Fuxiang Li}
	\email{fuxiangli@hnu.edu.cn}
	\affiliation{School of Physics and Electronics, Hunan University, Changsha 410082, China}
	
	\begin{abstract}
 Non-Hermitian physics provides an effective description of open and nonequilibrium systems and hosts many novel and intriguing phenomena such as exceptional points and non-Hermitian skin effect. 
Despite extensive theoretical and experimental studies, however, how to properly deal with the nonadiabatic dynamics in driven non-Hermitian quantum system is still under debate. Here, we develop a theoretical framework based on time-dependent  biorthogonal quantum formalism by redefining the associated state to obtain the gauge-independent transition probability,  and study the nonadiabatic dynamics of a linearly driven non-Hermitian system. In contrast to the normalization method that leads to a modified Kibble-Zurek scaling behavior, our approach predicts that the defect production at exceptional points exhibits power-law scaling behaviors conforming to the Kibble-Zurek mechanism. In the fast quench regime,  universal scaling behaviors are also found with respect to the initial quenching parameter, which can be explained by the impulse-adiabatic approximation.  Moreover, as trespassing the $\mathcal{PT}$-broken region,  the phenomenon of defect freezing, i.e., violation of adiabaticity, is observed. 
	 	\end{abstract}
	\maketitle

Non-Hermitian descriptions have been applied in a variety of physical systems, ranging from classical systems with gain or loss \cite{Makris2008,Klaiman2008,Malzard2015}, to open quantum systems \cite{ashida2020}, and many-body electronic system with interaction \cite{Yoshida2018,Shen2018,Nagai2020}. In non-Hermitian systems, many intriguing physics and interesting phenomena have been unveiled, such as exceptional points \cite{Hodaei2017,Bergholtz2021}, non-Hermitian skin effect \cite{Zhang2020,Zhang2022,Yao2018,Li2024}, tachyonic physics \cite{Lamata2007,Li2023}, and anomalous bulk-edge correspondence \cite{Lee2016,Yao2018,Song2019,Kunst2018,Xiong2018,Jin2019,Borgnia2020}. Non-Hermitian physics has been realized in diverse platforms, such as optical \cite{Zeuner2015}, microcavity \cite{Gao2015}, and acoustic systems \cite{Zhang2021}, electric circuits \cite{Wu2019}. 
In this work, we focus on the nonadiabatic dynamics of a non-Hermitian \cite{Xu2021,Pan2020,Lin2022,Liu2023,Geier2022,Kawabata2023,Yoshimura2020,McDonald2022,Zhai2022,agarwal2023,Agarwal2024,Roubeas2023} system and study the defect production due to a quench. It has been well known that, when a Hermitian system is slowly driven across the critical point, the produced defect density or excitation $n_{ex}$ scales as a universal power law with quenching time $\tau_Q$: 
\be\label{eq:KZM}
n_{ex} \sim \tau_Q^{-d\nu/(z\nu+1)} ,
\ee
where the exponent are determined by the static critical exponents $\nu$ and $z$.  This is the famous Kibble-Zurek (KZ) mechanism, a cornerstone of nonequilibrium physics,  which has been theoretically studied and experimentally verified in various platforms \cite{Deutschl2015,Maegochi2022,Anquez2016,Yi2020,Weiler2008,Ko2019,Du2023,Keesling2019,Ebadi2021,Lamporesi2013}. Whether the KZ scaling still holds at EPs in a non-Hermitian system is still under debate. In dissipative superconducting qubits governed by effective non-Hermitian Hamiltonians, it has been experimentally \cite{Doppler2016} observed a breakdown of adiabaticity, i.e., in the adibatic limit, the excitations are still nonzero. In theory, to study the dynamics of non-Hermitian system, a frequently adopted but problematic approach is to simply take the explicit normalization of time-dependent probabilities and observables   \cite{Turkeshi2023,Dora2019,Bacsi2021,sticlet2022,Dora2020,longs2019,Lu2024,Karin2024}. Using this approach, recent work predicted a recovery of adiabaticity and a modified KZ scaling  when a non-Hermitian system is quenched across EPs \cite{Dora2019,Xiao2021}
\be\label{eq:gKZM}
n_{ex}^m \sim\tau_Q^{-(d+z)\nu/(z\nu+1)} ,
\ee 
More recently, in Ref.~\cite{Sim2023}, a formulation based on the metric framework was proposed, in which the Hilbert space is nonstationary and endowed with a nontrivial time-dependent “metric”, and therefore, the norm of the wavefunction is conserved \cite{Mostafazadeh2020,Geyer2008,Mostafazadeh2018}. This approach predicted that $\mathcal{PT}$-broken time evolution leads to defect freezing and hence the violation of adiabaticity. These conflicting results indicate that a natural and consistent treatment is necessitated  for the nonadiabatic dynamics of a general time-dependent non-Hermitian Hamiltonian.    

Here, we address these issues by developing a new theoretical framework  based on the  biorthogonal quantum mechanics \cite{Kunst2018,Brody2014,Edvardsson2019,Edvardsson2020} for the study of nonadiabatic dynamics of general non-Hermitian systems. By introducing a new associated state to obtain the gauge-invariant transition probability, our theory not only restores the traditional KZ scaling, but also recovers the defect freezing phenomenon.  
Specifically, we study the nonadiabatic dynamics of a linearly driven non-Hermitian Su-Schrieffer-Heeger model, as an example. In contrast to the normalization method that leads to a modified Kibble-Zurek scaling behavior (\ref{eq:gKZM}), our approach predicts that the defect production at exceptional points  exhibits power-law scaling behaviors with respect to quench time, with the scaling exponent conforming to the KZ mechanism (\ref{eq:KZM}). In the fast quench regime, the excitation density and critical quench time are found to scale with the initial quench parameter, which can be explained by the impulse-adiabatic approximation.  Moreover, when the system trespasses the $\mathcal{PT}$-symmetry-broken region,  our theory predicts that the excitation does not vanish even in the adiabatic limit, and thus leads to the violation of nonadiabaticity.

{\it Biorthogonal quantum mechanics.} For a general non-Hermitian Hamiltonian $H\neq H^{\dagger}$, one should define both the right and left eigenstates: 
\be
H\vert u_n\rangle=E_n\vert u_n\rangle, \quad\quad \langle u_n\vert H^{\dagger}=E_n^{\ast}\langle u_n\vert, 
\ee
and they  satisfy the completeness relation $\sum_{n}|\tilde{u}_n\rangle\langle u_n|=\mathds{1}$ and the biorthonormal relation $\langle \tilde{u}_m|u_n\rangle=\delta_{m,n}$. 
The right eigenstates $\{|u_n\rangle\}$ of $H$ are in general not orthogonal, 
 i.e., $\langle u_n|u_m\rangle\neq0$ if $n\neq m$,  which causes serious problem in the probabilistic interpretation in non-Hermitian quantum mechanics. 
 To reconcile these apparent contradictions, the so-called associated state and the redefinition of the inner product have been introduced \cite{Brody2014}. 
For an arbitrary state $|\psi\ra$ expressed in terms of the linear superposition of right eigenstates, one can define its associated state $|\tilde{\psi}\rangle$ in the following way:
\be
|\psi\ra = \sum_n c_n |u_n\ra, ~~|\tilde{\psi}\rangle=\sum_n c_n |\tilde{u}_n \ra. \label{eq:associated1}
\ee
The transition probability between the $|\psi\rangle$ and the eigenstate $|u_n\rangle$ in the biorthogonal bases is defined as: 
\be
p_n=\frac{\langle\tilde{\psi}|u_n\rangle\langle \tilde{u}_n|\psi\rangle}{\langle\tilde{\psi}|\psi\rangle\langle \tilde{u}_n|u_n\rangle},
\label{eq:pn}
\ee
satisfying the normalization condition $\sum_{n}p_n=1$.  

		However, the associated state in Eq.(\ref{eq:associated1}) is not well-defined. It contains a gauge  ambiguity that would lead to an ill-defined probability (\ref{eq:pn}). 
  Without loss of generality, we  rescale each basis vector $|u_n \ra$ by a factor $a_n$: 
		\be
		|u_n'\rangle=\frac{1}{a_n} |u_n\rangle,~~ |\tilde{u}_n'\rangle=a_n|\tilde{u}_n\rangle. \label{eq:scale}
		\ee
 Obviously, the new basis is still orthogonal and normalized according to the biorthogonal quantum mechanics:   $\langle \tilde{u}'_m|u_n'\rangle= \delta_{m,n}$. In this new basis, the same quantum state $|\psi\ra$ can be expanded with  new coefficients $c_n'$, and a new associated state $|\tilde{\psi}'\ra$ is obtained according to (\ref{eq:associated1}):
		\be
		|\psi\rangle =\sum_nc_n'|u_n'\rangle,~~~|\tilde{\psi}'\rangle=\sum_nc_n'|\tilde{u}_n'\rangle
		\ee
Now we have $c'_n= a_n c_n$, and one finds that, returning to the original basis, the new associated state is different from the original associated state $|\tilde{\psi} \ra$: 
\be
|\tilde{\psi}' \ra = \sum c_n a_n^2 |\tilde{u}_n \ra \neq |\tilde{\psi} \ra
\ee
Correspondingly, one can easily check that the transition probability $p_n$ is also changed after this rescaling (\ref{eq:scale}). 

To eliminate this ambiguity, in this paper, we provide a new framework for the associated state by fixing the gauge freedom so that the associated state is expanded in terms of normalized left eigenstates:
\be
|\tilde{\psi}\rangle=\sum_n c_n\frac{|\tilde{u}_n\rangle}{\langle \tilde{u}_n|\tilde{u}_n\rangle}. \label{eq:assoct}
\ee
It is ready to check that the transition probability (\ref{eq:pn}) is now a well defined quantity and remains unchanged under rescaling of bases (\ref{eq:scale}).

We  next  discuss  the time evolution of an arbitrary initial state $|\psi(0)\rangle$ in a time-dependent driven non-Hermitian system. After solving  the time-dependent Schr$\mathsf{\ddot{o}}$dinger equation $i\frac{d}{dt}|\psi(t)\rangle=H(t)|\psi(t)\rangle$, one can express the state vector $|\psi(t)\ra$ in terms of the instantaneous eigenstates $|u_n(t)\ra$ of time-dependent Hamiltonian $H(t)$, $|\psi(t)\ra = \sum_n c_n(t) |u_n(t) \ra$. Here, $c_n=\langle \tilde{u}_n(t)|\psi(t)\rangle$, and $H(t) |u_n(t) \ra = E_n(t) |u_n(t) \ra $.   We can thus define the time-dependent version of associated state in terms of the instantaneous eigenstates $|\tilde{u} _n(t)\rangle$ of  $H^{\dagger}(t)$: 
\be
|\tilde{\psi}(t)\rangle=\sum_{n}c_n\frac{|\tilde{u} _n(t)\rangle}{\langle\tilde{u} _n(t)|\tilde{u} _n(t)\rangle}. 
\ee
The transition probability between the $|\psi(t)\rangle$ and $|u_n(t)\rangle$ now can be written as :
 \be\label{eq:pk1}
 p_{n}(t)=\frac{\langle\tilde{\psi}(t)|u_n(t)\rangle\langle\tilde{u}_n(t)|\psi(t)\rangle}{\langle\tilde{\psi}(t)|\psi(t)\rangle\langle\tilde{u}_n(t)|u_n(t)\rangle}.
 \ee
 With the definitions of transition probability $p_n(t)$, one can calculate the excitations at the end of a quench  by letting $|\psi(0)\rangle$ be the ground state of the initial Hamiltonian $H(0)$ and $|u_n(t_f)\rangle$ the excited state of final Hamiltonian $H(t_f)$. 
 
  \begin{figure*}[t]
	\centering
	\includegraphics[width=1\textwidth]{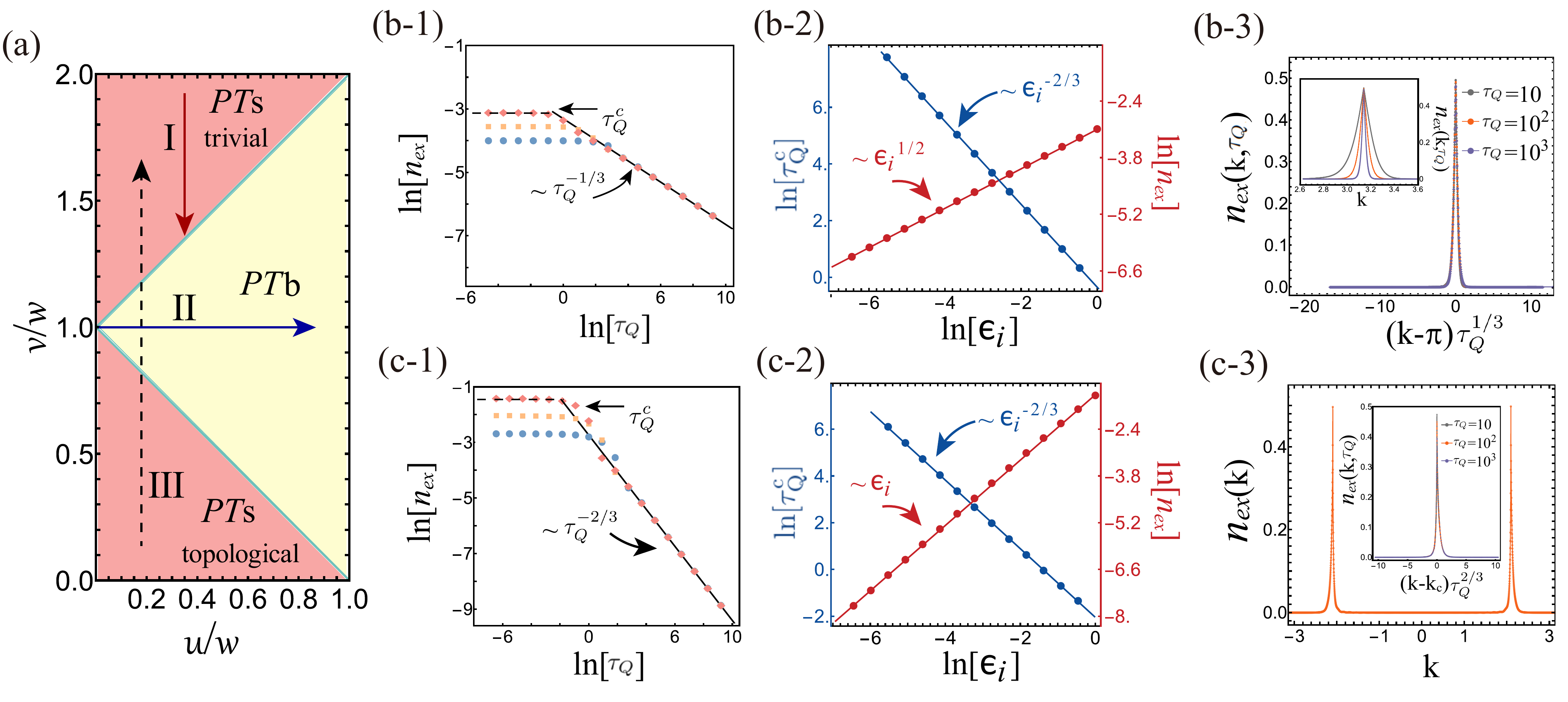}\\
	\caption{(a) Phase diagram of the non-Hermitian SSH model. Three phases are identified: $\mathcal{PT}$-symmetric ($\mathcal{PT}s$) topological, $\mathcal{PT}s$ trivial and $\mathcal{PT}$-broken ($\mathcal{PT}b$) phase. The different quench protocols are denoted by the different arrows. The scaling behaviors of excitations during the $\mathcal{PT}$-symmetric quench (b) and  full non-Hermitian quench (c).  (1) The density of excitation vs quench time $\tau_Q$ for three different initial values of quenching parameters.  In the slow quenches regime, the density is consistent with the KZM scaling. In thee fast quench regime, the density of excitation saturates with  saturation value depending on initial values of quenching parameter. (2) The saturation value of excitation density $n_{ex}$ and the critical quench time $\tau_Q^c$ exhibit universal power-law scaling with the dimensionless distance  $\epsilon_i$ from the exceptional point. (b-3) Curves of  momentum resolved excitations density $n_{ex}(k,\tau_Q)$ versus $k$ collapse onto a single line after rescaling (\ref{eq:hres}) $z=1$ and $\nu=1/2$. The inset shows the curves before rescaling. (c-3) The distribution of  $n_{ex}(k,\tau_Q)$ at the end of the quench with $u(t_f)=1$ for the non-hermitian drive, which exhibits two peaks at $k_c=\pm \frac{2\pi}{3}$, respectively.  The rescaling results according to (\ref{eq:hres}) with $z=1/2$ and $\nu=1$ is shown in the inset. For clarity, we only show the results near the $k_c=\frac{2\pi}{3}$. }\label{fig1}.
\end{figure*}

{\it Model.} In this paper, we consider the non-Hermitian Su-Schrieffer-Heeger (SSH) model \cite{Lieu2018,Gong2018,Chang2020} with the $\mathcal{PT}$-symmetry. The Hamiltonian defined in momentum space is given by
\be\label{eq:hk}
H_k=(v+w\cos k)\sigma_x+ w \sin k\sigma_y+i u\sigma_z, \label{eq:model}
\ee
where $\sigma_x, \sigma_y,\sigma_z$ are Pauli matrices and the parameters  $w, v$ and $u\in\mathbb{R}$ are assumed to be positive without loss of generality. $\mathcal{PT}$-symmetry is realized in the model by the operator $\mathcal{P}=\sigma_z$ and $\mathcal{T}=-i \sigma_yK$ where $K$ is complex conjugation, such that $[H_k,\mathcal{PT}]=0$. At the EP, the $\mathcal{PT}$-symmetry is spontaneously broken and the states are no longer eigenstates of the $\mathcal{PT}$ operator. The instantaneous eigenvalues of Eq.~(\ref{eq:hk}) are given by $E_{k,\pm}=\pm \sqrt{|w-v|^2+2wv(1+\cos k)-u^2}$. When $|w-v|>u$, $H_k$ has real eigenvalues for each $k$ ($\mathcal{PT}$-symmetry regime). In the regime with $u>|w-v|$ ($\mathcal{PT}$-broken regime), however, $H_k$ has complex eigenvalue for certain $k$. It is worth noting that in the $\mathcal{PT}$-symmetry regime there are two different topological phases. For the region with $w-v>u$, the system is in the topological phase and there exists a pair of protected edge modes under open boundary conditions, while in the $w-v<-u$ region, the system is topologically trivial and does not support edge modes. 

The phase diagram is plotted in Fig.~\ref{fig1}(a),  which shows that three different quench protocols are considered: (1) $\mathcal{PT}$-symmetric quench with varying $v(t)$; (2) full non-Hermitian quench with varying $u(t)$; and (3) quench through the $\mathcal{PT}$-broken region. 

For the convenience of the description, the transition probability $p_k$ for a given mode $k$ in our model (\ref{eq:model}) can be denoted as $n_{ex}(k)$ referred to as momentum resolved excitation density. The density of excitations is  given by $
 n_{ex}=\frac{1}{L} \sum_{k}n_{ex}(k) $ with  $L$ being the  system length.

{\it $\mathcal{PT}$-symmetric quench.} Now we consider a fully $\mathcal{PT}$-symmetric ramp during which the instantaneous spectrum is always real. 
In this quench protocol, we fix the parameters $w=1$, $u=1/2$ and the EP is at $v_c=3/2$. We linearly vary the $v$  as $\label{eq:quench1}
v(t)-v_c=v_i(1-\frac{t}{\tau_Q})$  during the time interval $t\in [0,\tau_Q)$.  Since the instantaneous eigenstates coalesce with each other at  EP and one cannot define the transition probability  using Eq.~(\ref{eq:pk1}), the ramp does not end precisely at EP but very close to it, to ensure that the end point is within the freeze-out  zone. 

The numerical results for different $v_i$ are shown in Fig.~\ref{fig1}(b-1), which shows that, in the adibatic limit with large $\tau_Q$, the density of excitations produced after the $\mathcal{PT}$-symmetric quench exhibits a power-law behavior with the quench time $\tau_Q$: $ n_{ex}\propto \tau_Q^{-1/3}$. Note that, along the quench trajectory, the energy gap scales as   $\Delta\propto\sqrt{v-v_c}$ for $v\sim v_c$, while near the critical point,  the energy spectrum scales as $E_{\pm}\propto\pm|k-\pi|$. One thus has  $z\nu=1/2$  and  $z=1$, leaving us with $\nu=1/2$.   Therefore, the exponent $1/3$  satisfies the KZM scaling (\ref{eq:KZM}) rather than the modified one (\ref{eq:gKZM}).

In the fast quench regime with small $\tau_Q$, one observes that the excitation density approaches to a plateau with saturated value independent of the quench time $\tau_Q$. In the fast quench regime, a generalized KZ scaling has  been demonstrated that the defect density exhibits a universal power-law scaling with the final of the control parameter \cite{Zeng2023,Yang2023, Grandi2010}. In our case, since we are considering the quench of a control parameter $\lambda$ from initial value $\lambda_i$ to the critical point $\lambda_c$ \cite{ Grandi2010}, the response of the system would depend only on the initial value $\lambda_i$. For fast quenches with relaxation time $\tau(\lambda_i)>\tau[\lambda(\hat{t})]$, the freeze-out time is given by $\hat{t}\sim \tau(\lambda_i)\propto \epsilon_i^{-z\nu}$ with  $\epsilon_i=|\lambda_i-\lambda_c|/\lambda_c$ the reduced distance parameter. 
The defect density is determined by the correlation length $\xi(\lambda_i)$ at initial value of the control parameter $\lambda_i$: 
\be\label{eq:dv}
n_{ex}\sim\frac{1}{\xi(\lambda_i)^d}\propto\epsilon_i^{d\nu},
\ee
which is universal and independent of the quench time. Moreover, a critical quench time $\tau_Q^{c}$ separating the KZ scaling regime to the saturation regime, is defined by equating the time at which the quench ends at the critical point, $t^c=\tau_Q^c |\epsilon_i|$, to the relaxation time at $\epsilon_i$. The condition  $\tau_Q^{c} |\epsilon_i|=\tau_0/\vert\epsilon_i|^{z\nu}$ leads to the scaling of  critical quench time $\tau_Q^c$ with the initial control parameter $\epsilon_i$: 
\be\label{eq:1+zv}
\tau_Q^{c}\propto\epsilon_i^{-(1+z\nu)}.
\ee
  In Fig.~\ref{fig1}(b-2), we show that both the saturated value of defect density and the critical quench time $\tau_Q^{c}$  exhibit universal power-law scalings with the initial value $\epsilon_i$. The exponents are, respectively $1/2$ and $3/2$,  in agreement with  prediction  (\ref{eq:dv}) and (\ref{eq:1+zv}) .

Fig.~\ref{fig1}(b-3), we also study the scaling behaviour of the momentum resolved excitation density, which satisfies $ n_{ex}(k,\tau_Q)=F[(k-\pi)\tau_Q^{1/3}]$ with $F(x)$ a nouniversal scaling function. This relation is in agreement with the scaling behavior in a Hermitian system, in which 
\be\label{eq:hres}
 n_{ex}(k,\tau_Q)=F(k^z\tau_Q^{z\nu/(z\nu+1)}).
 \ee
This result is in sharp contrast with Ref. \cite{Dora2019} which proposed that in non-Hermitian systems, the induced defect vanishes as $\tau_Q^{-(d+z)\nu/(z\nu+1)}$ due to nonorthogonality of wavefunctions. 


 \begin{figure}[t]
	\centering
	\includegraphics[width=0.45\textwidth]{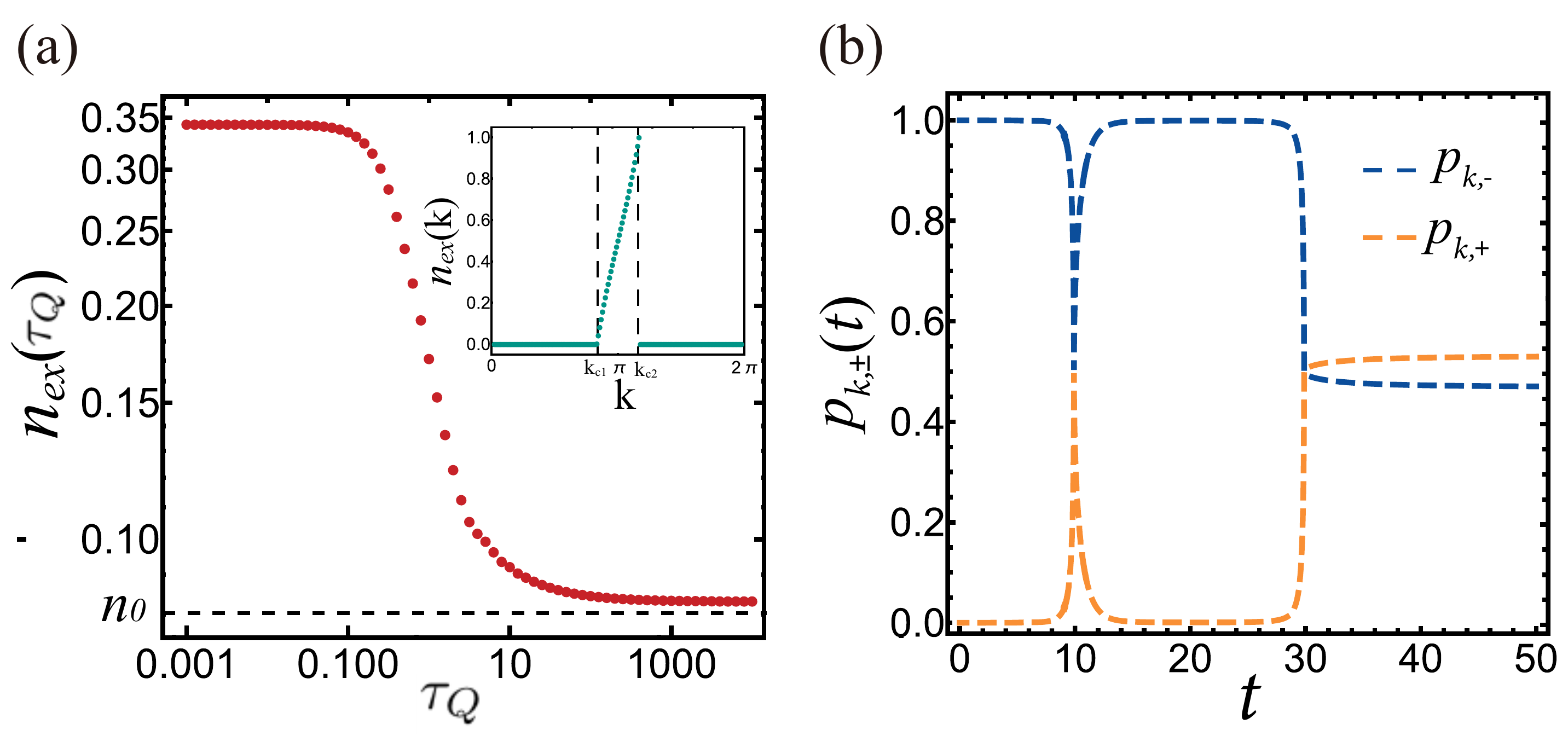}\\
	\caption{Quench through the $\mathcal{PT}$-symmetric broken regime. (a) The density of excitations in function of $\tau_Q$. In the adiabatic limit $\tau_Q\rar \infty$, the density $n_{ex}$ tends to a nonzero constant value $n_0$, which suggests the violation of quantum adiabaticity. The inset shows the excitations contribution from the $\mathcal{PT}$-broken modes . (b) For a $\mathcal{PT}$-broken $k$ mode, transition probability $p_{k, \pm}(t)$ as a function of time $t$. In the adiabatic limit, the finial value is independent of the initial state. } \label{fig2}
\end{figure}

{\it Full non-Hermitian drive}. A full non-Hermitian drive is realized for $v=w=1, u(t)=u_c-u_i(1-t/\tau_Q)$ with critical point at $u_c=1$ during $t\in[0,\tau_Q)$, which represents the non-Hermitian Kibble-Zurek problem and is equivalent to quenching the imaginary tachyon mass \cite{Lee2019}. 

During the time evolution, the instantaneous eigenenergies remain real for the modes $|k| < k_c$, while for modes $|k|>k_c$, they experience a $\mathcal{PT}$ symmetry breaking, and thus change from real to imaginary. Here, the critical momentum $k_c=\pm2\pi/3$, for which the EP is located at the end of quench (See SM for details). 
In the numerical calculation, the ground state is chosen to be the least-dissipative eigenstate with the largest imaginary eigenvalue \cite{Tong2023}. Therefore, in the adiabatic limit, the excitations are dominated by the modes near $k_c$. 

In the vicinity of EPs, the spectrum scales as  $E_{\pm}(|k|\lesssim |k_c|)=\pm\sqrt{\sin |k_c|}\sqrt{|k_c|-|k|}$ in the $\mathcal{PT}$-symmetric regime, and as $E_{\pm}(|k|\gtrsim |k_c|)=\pm i\sqrt{\sin |k_c|}\sqrt{|k|-|k_c|}$ in the $\mathcal{PT}$-symmetry broken regime. Moreover, near the end of the quech, the energy gap scales as $\Delta\propto|u-u_c|^{1/2}$. One therefore can determine the exponents as $z=1/2$ and $\nu=1$. 



Fig.~\ref{fig1}(c-1) shows the scaling of excitation density with quench time as $n_{ex}\propto \tau_Q^{-2/3}$ in the slow quench regime, which conforms to the prediction of KZ scaling (\ref{eq:KZM}).  Fig.~\ref{fig1}(c-2) shows the scalings of saturated value of excitation density $n_{ex}$ and critical quench time $\tau_Q^{c}$  with initial value $\epsilon_i$ as $n_{ex}\propto\epsilon_i$ and $\tau_Q\sim\epsilon_i$ in the fast quench regime. Here,  $\epsilon_i = (u(\tau_Q)-u(0))/u(\tau_Q) = (u_i -u_c)/u_c$.  Bot exponents conform to the predictions of fast quench KZ scaling (\ref{eq:dv}) and (\ref{eq:1+zv}). The momentum-resolved excitation is also plotted in Fig.~\ref{fig1}(c-3),  which exhibits two peaks at  $k_c=\pm 2\pi/3$.  The inset of  Fig.~\ref{fig1}(c-3) shows that the plots for different quench time $\tau_Q$ collapse to a common scaling function according to $n_{ex}(k,\tau_Q)=F_{nh}[(k-k_c)^{1/2}\tau_Q^{1/3}]$ as predicted by (\ref{eq:hres}). 




{\it Quench through the $\mathcal{PT}$-broken region.}
For the  quench protocol III of Fig.~\ref{fig1}(a), we fix the parameters $w=1$, $u=1/2$ and vary the $v$ as  $v(t)=1+t/\tau_Q$ during the time interval $t\in [-\tau_Q,\tau_Q]$. 
This quench crosses the phase boundary twice. The numerical results of the excitation density as a function of $\tau_Q$ are plotted in Fig.~\ref{fig2}(a). In contrast to the previous two quench protocols, Fig.~\ref{fig2}(a) shows that, even in the adiabatic limit, the excitation does not vanish,  indicating the violation of quantum adiabaticity, i.e., defect freezing.  The inset of Fig.~\ref{fig2}(a) further shows that the nonzero excitations come from the contribution of the $\mathcal{PT}$-broken modes.  The breakdown of adiabaticity was also observed experimentally in dissipative superconducting qubits governed by effective non-Hermitian Hamiltonians \cite{Doppler2016} and captured by quantum metric framework \cite{Sim2023}. 

Fig.~\ref{fig2}(b) illustrates the time-dependent transition probability $p_k(t)$ for a specific $\mathcal{PT}$-broken mode $k$ during the time-evolution. At the beginning of the evolution, the energies of the two-level system  are real, and the energy gap is large enough so that the initial ground state evolve adiabatically. When the Hamiltonian approaches an EP, the dynamics cannot be adiabatic and  the system
gets excited.  After crossing the EP, the eigenvalues of the time-dependent Hamiltonian  are complex conjugate and there is an exponentially growing state and an exponentially decaying state. And the component corresponding to the decaying state in the time-evolved state gradually vanishes  during subsequent evolution. Thus the system decays to the ground state (least-dissipative instantaneous eigenstate) in the adiabatic limit. As approaching the second EP, the system gets excited again, which results nonzero excitations \cite{longs2019,pan2024}.


{\it Conclusion.~} We have developed a natural and consistent theoretical framework for the study of nonadiabatic dynamics of general non-Hermitian Hamiltonians. We have introduced a new definition of associated state, and under this definition, the transition probability is gauge-invariant and thus is physically reasonable. We apply this approach on the linearly driven non-Hermitian model, and find that not only the KZ scaling is restored, but also the defect freezing is recovered for different quench protocols. Our approach resolves the conflicting  results in the literature. Our work paves a way to explore the rich nonequilibrium phenomena in non-Hermitian systems.

\bibliographystyle{apsrev4-2}
%


\end{document}